\documentclass{ws-jai}
\usepackage[flushleft]{threeparttable}
\begin{document}

\catchline{}{}{}{}{} 

\markboth{G. Comoretto et al.}{Signal Processing Firmware for the LFAA}

\title{The Signal Processing Firmware for the Low Frequency Aperture Array}

\author{Gianni Comoretto$^1$,
Riccardo Chiello$^2$, 
Matt Roberts$^3$,
Rob Halsall$^3$,
Kristian Zarb Adami$^{2,5}$,
Monica Alderighi$^4$,
Amin Aminaei$^2$,
Jeremy Baker$^3$,
Carolina Belli$^1$,
Simone Chiarucci$^1$,
Sergio D'Angelo$^4$,
Andrea De Marco$^5$,
Gabriele Dalle Mura$^6$,
Alessio Magro$^5$,
Andrea Mattana$^7$,
Jader Monari$^7$,
Giovanni Naldi$^7$,
Sandro Pastore$^8$,
Federico Perini$^7$,
Marco Poloni$^7$,
Giuseppe Pupillo$^7$,
Simone Rusticelli$^7$,
Marco Schiaffino$^7$,
Francesco Schillir\`o$^9$,
and Emanuele Zaccaro$^6$
}

\address{
$^1$Istituto Nazionale di Astrofisica - Osservatorio Astrofisico di Arcetri,
Largo E. Fermi, 5, 50125 Firenze, Italy, comore@arcetri.astro.it\\
$^2$University of Oxford, Denys Wilkinson Building, Oxford, OX1 3RH, 
United Kingdom, riccardo.chiello@physics.ox.ac.uk \\
$^3$Science \& Technology Facilities Council, Rutherford Appleton Laboratory, 
Harwell Campus, Didcot, OX11 0QX, United Kingdom, matt.roberts@stfc.ac.uk, rob.halsall@stfc.ac.uk \\ 
$^4$Istituto Nazionale di Astrofisica - Istituto di Astrofisica Spaziale
e Fisica Cosmica, Via E. Bassini 15, I-20133 Milano, Italy\\
$^5$Institute of Space Sciences and Astronomy, University of Malta, 
Msida, Malta\\
$^6$ Campera ES, Via Mario Giuntini 13, 56023, Navacchio, Pisa, Italy \\
$^7$ Istituto Nazionale di Astrofisica - Istituto di Radioastronomia,
Via P. Gobetti, 101, 40129 Bologna, Italy\\
$^8$ Sanitas EG, Viale F. Restelli, 3, 20124 Milan, Italy \\
$^9$ Istituto Nazionale di Astrofisica - Osservatorio Astrofisico di Catania,
Via S.Sofia 78, 95123 Catania, Italy\\
}

\maketitle

\begin{history}
\received{(to be inserted by publisher)};
\revised{(to be inserted by publisher)};
\accepted{(to be inserted by publisher)};
\end{history}

\begin{abstract}
The signal processing firmware that has been
developed for the Low Frequency Aperture Array component of the Square
Kilometre Array is described. The firmware is implemented on a dual FPGA board,
that is capable of processing the streams from 16 dual polarization
antennas. Data processing includes channelization of the sampled
data for each antenna, correction for instrumental response and for
geometric delays and formation of one or more beams by combining the
aligned streams. The channelizer uses an oversampling polyphase 
filterbank architecture, allowing a frequency continuous processing 
of the input signal without discontinuities between spectral channels. 
Each board processes the streams from 16 antennas,
as part of larger beamforming system, linked by standard
Ethernet interconnections. There are envisaged to be 8192
of these signal processing platforms in the first phase of the Square
Kilometre array so particular attention has been devoted to ensure the
design is low cost and low power. 
\end{abstract}


\section{Introduction}

\begin{figure}[htb]
\begin{center}
\includegraphics[width=0.5\textwidth]{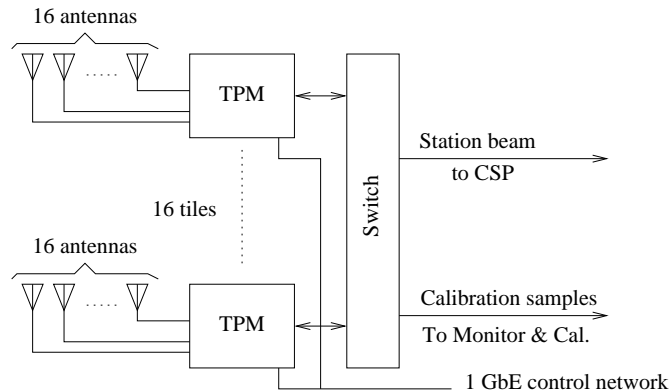}
\end{center}
\caption{
Architecture of the Low Frequency Aperture Array station. 
Groups of 16 antennas form a telescope tile. 16 tiles form a station, that
produce a station beam to be processed in the central signal processor
}
\label{fig:arch}
\end{figure}

The Low Frequency Aperture Array (LFAA) \cite{faulkner01}
component of the Square Kilometre
Array (SKA) will consist, in SKA Phase 1, of $2^{17}$ log-periodic 
dipole antennas (SKALA). 
The current
architecture, shown in Figure \ref{fig:arch}, envisages the transport of
the Radio-Frequency (RF) bandwidth of 300~MHz over fiber to a Central
Processing Facility (CPF). Once inside the building these antenna
signals are digitized, channelized and beamformed together to form
logical stations which are an aggregation of 256 antennas.  The LFAA is
currently designed to produce 512 of these logical stations which can
be flexibly configured by programming the signal processing platform to
send its traffic across a highly configurable network.

The antennas are grouped in tiles of 16 antennas each. Each tile is
processed in a Tile Processing Module (TPM), and 16 TPMs are connected
together in a flexible way to form a station, using a general purpose high
speed Ethernet interconnect. This distributed beamformer architecture 
allows to dynamically reconfigure the antennas composing each station. 

Signal processing inside a LFAA station includes a first channelization 
stage, corrections for cable length mismatch, geometric delay,
receiver amplitude and phase response, and atmospheric and polarization
calibrations.  Beamformed samples are then organized into frames and
sent to the SKA Central Signal Processor (CSP).  The CSP provides a
second channelization stage, correlation, beamforming for pulsar search
and timing.

The two stage oversampling filterbank architecture allows for a very
large number of spectral channels, avoiding the spectral holes present in a 
conventional two stage channelizer.
In previous instruments a "twice oversampled" architecture
was sometimes used, at the cost of doubling the amount of data to
be transferred between the two stages \cite{comore11}, but this
was not acceptable in an instrument as large as SKA. An oversampled
filterbank has been used in the correlator for the ALMA interferometer
\cite{escoffier07}, implemented as a set of 32 tunable digital
down-converters, but this approach is not scalable to the number of
channels required here.

This paper describes the signal processing algorithms, procedures
and instrumental effects, in the context of the SKA Low telescope signal
processing chain, while the TPM hardware is described in detail elsewhere 
\cite{naldi16}. Section \ref{sec:proc} provides an overview of the
LFAA signal processing, and subsequent sections describe in detail the
ADC conversion and synchronization (section \ref{sec:adc}), the channelizer
(section \ref{sec:pfb}), the beamformer (section \ref{sec:bmf}),
the interface for data transport between tiles and towards the CSP
(section \ref{sec:spead}), and diagnostic and calibration functions
(section \ref{sec:diag}). Some tests and implementation 
details are shown in section \ref{sec:results}.

\section{LFAA signal processing}
\label{sec:proc}

The LFAA signal processing subsystem is responsible for the formation 
of one or more beams from each station.
Station beams are channelized to a
resolution of 781.25~kHz, covering the frequency range of 50
to 350 MHz.  Each beam is composed of the signals from 256 antennas,
coherently added along one or more directions in the sky. Up to 8 beams
may be formed, with a total aggregate bandwidth of 300 MHz (e.g. 6 beams
of 50 MHz each).  Beams must be capable of following a source moving in
the sky at up to several times the sidereal rate.

Both the SKA Low and SKA Mid telescopes adopt a 2-stage  channelization
scheme, with a first stage oversampled polyphase filterbank followed by
a second stage critically sampled filterbank implemented in the CSP.
Oversampling must not exceed 20\%, in order to limit the amount of data
transferred to the CSP.  The first stage LFAA channelizer provides a
flat and alias free frequency response on a region equal to the channel
separation. The slightly larger bandwidth (oversampling) is used to
accommodate the portion of the band affected by the filter transition
zone and aliasing of the adjacent channels.  The alias free region
in each frequency channel is further channelized in the CSP, with a
variable resolution ranging from 226 Hz to about 5.4 kHz, while the
portion affected by aliasing and filter edges is discarded. These {\em
fine channels} are then stitched together to form a continuous uniform
spectrum at the desired final frequency resolution.  In-band ripple is
limited to $\pm0.2$~dB, minimum stop-band rejection must exceed 60 dB,
with more than 80 dB as a goal for most of the stop-band region.

The TPM must deal with several instrumental effects. Cable length mismatches 
among antennas must be corrected in a static way, for pointing to zenith. 
Receiver and antenna gain and phase frequency responses must be corrected 
with a frequency resolution of one spectral channel, and corrections are 
updated every 10 minutes during the observation, without stopping it. 
Polarization impurities and polarization rotation must be individually
corrected for each beam. The signal must be equalized in frequency,
to accommodate for the frequency dependence of the observed signal
spectral density, in order to keep the amplitude of the signals sent to
the CSP within $\pm3$~dB. 

Beside the beamformed data, the TPM must produce some diagnostic
quantities for the calibration subsystem. 
The station calibration algorithm requires a continuous stream of
channelized samples from one selected spectral channel.
The total power and a coarse frequency spectrum, computed over a limited
integration time, must also be provided for each input signal.

The TPM is connected to the local monitor and control (LMC) 
subsystem via a 1Gb
Ethernet interface. A control software framework has been 
developed \cite{magro16}, including a low level communication 
layer that uses a symbol table embedded in the firmware, and
higher level functions that interface to a TANGO control 
middleware \cite{tango15}. Most of the parameters are set before the actual
observation, but all calibration quantities (including beamforming
parameters) must be changed dynamically at predefined times, without 
stopping the observation.

Beamformed data are sent to the CSP using a 40 Gb high 
speed Ethernet interconnect. The same network is also used to send the
channelized samples to the LFAA calibration subsystem.

The TPM is responsible of time stamping of the samples
from the Analog to Digital Converter (ADC). Each TPM is 
synchronized to Universal Time Coordinated (UTC) using a 
peak-per-second (PPS) pulse and a reference high precision clock, 
with UTC time distributed via the control interface.

The LFAA antennas operate in a quiet Radio Frequency Interference (RFI)
region, but the telescope is anyway very sensitive 
to RFI.  The channelization subsystem must be able to cope with
RFI, limiting the effects to the affected time and frequency regions.

\subsection{Tile Processing Module architecture}

The LFAA station beamformer structure is based on a frequency domain
beamforming architecture. Data streams from the individual antennas are
channelized with a channel spacing of 781.25~KHz, and delayed
in the frequency domain by applying a dynamic phase correction to
each individual channel. Each signal is also corrected using a static
instrumental gain, phase and polarization calibrations, updated on a
timescale of a few minutes. It is possible to select multiple regions
in the processed band, and/or to generate multiple beams of the same
or different spectral regions. The channelization process is used both
to allow for frequency dependent calibration, 
for frequency domain beamforming, and to provide
the first stage of a two-stage channelizer.

Processing is performed in groups (tiles) of 16 antennas in a TPM. TPMs
are connected together in a flexible way using the 40 Gb high speed
network, with non-blocking network switches.  Stations can then be configured
dynamically as arbitrary groups of tiles. The same network is used to
transfer samples from individual antennas to the control, calibration and
monitor subsystem, and beamformed samples to the Central Signal Processor.
Data to be sent to the CSP is packed into frames of 2048 samples for
one frequency channel, two polarizations. This requires a corner turner
operation from the channelizer output (frames of one time sample, all
channels) to the CSP frames (one channel, many consecutive time samples).

\begin{figure}
\begin{center}
\includegraphics[width=0.9\textwidth]{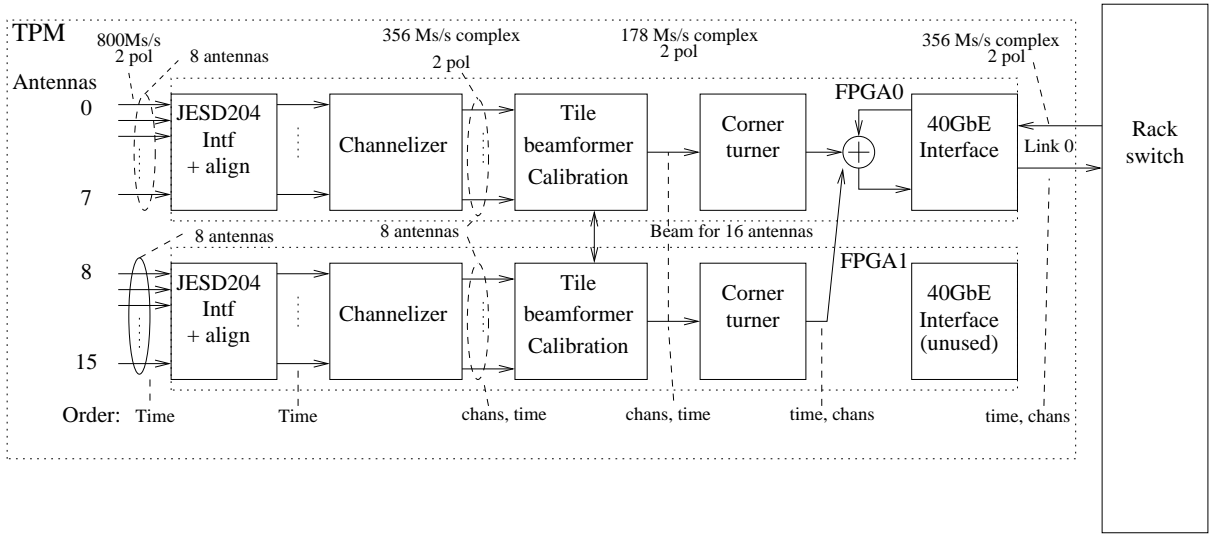}
\end{center}
\caption{
Signal processing chain of a tile processing module
}
\label{fig:tpm}
\end{figure}

The firmware has been optimized for the Italian TPM, 
described in detail in \cite{naldi16}. It includes 16 
dual channel ADCs connected in groups of 8 to two separate 
Xilinx Kintex Ultrascale Field Programmable Gate Arrays (FPGA), each
processes both polarizations for 8 antennas (Figure \ref{fig:tpm}).
The two FPGAs are interconnected by a parallel high speed bus, with 
an aggregate bandwidth of 22 Gbps in each direction. Each FPGA has 
an external memory bank, and a  Quad Small Form-factor Pluggable 
(QSFP) connector for up to two 40~Gb Ethernet interfaces, but only
one QSFP will be used in the full scale SKA system for cost and 
power considerations.

The structure is adapted to a time multiplexed data stream, with 4
samples processed in parallel at each FPGA clock cycle.  
After channelization and beamforming, odd and even frequency channels for
the whole tile beam are processed separately in the two FPGAs. Channelized
data is stored in an external local memory, that is used for the corner
turner function. Station beamforming is performed by daisy-chaining
the TPMs. The first TPM in the chain retrieves from memory the samples
corresponding to a packet to be sent to the CSP, and sends it to the
next one, that in turn retrieves the corresponding samples, adds them
to the incoming packet and forwards it down the chain. The last TPM then
reformats the packet and sends it to the CSP. 

Data rates are also shown in figure \ref{fig:tpm}. Each ADC generates
800 Mbyte/s of 8 bit samples, for an aggregate data rate of 204 Gbps. 
The channelizer selects a bandwidth of 300 MHz, oversampled to 356 
Msample/s, with resolution increased to 12+12 bit complex samples. 
The aggregated data rate after the channelizer is then 274 Gbps, 
that is reduced by a factor
of 16 (one beam every 16 antennas, 17 Gbps) in the tile beamformer. 
Half of this information is exchanged among the two FPGAs. 
The traveling sum on the 40 Gb interface uses 16+16 bit complex samples, 
for a data rate, including packet overhead, of 23.2 Gbps. The final 
beamformed data to the CSP is represented with 8+8 complex samples, 
with a data rate of 11.6 Gbps.

The calibration subsystem requires a subset of the channelized samples
for one channel, all antennas and polarizations, represented with 8+8 
bit complex samples. The corresponding aggregated data rate is about 
0.5 Gbps. 

\section{ADC interface}
\label{sec:adc}

The TPM hosts 16 dual-input AD9680 digitizers sampling 16 dual polarization
antennas at 800 megasample/second (MSPS). Both the ADC and the signal 
processing chain can operate in the second Nyquist zone, and with slightly 
different digitization rates, but this possibility is not used in SKA1.
Digitized data are transferred towards 
two Xilinx Kintex UltraScale XCKU040 FPGAs using fast serial interfaces
exploiting JESD204B Subclass 1, which guarantees deterministic latency. 

\subsection{JESD interface synchronization}

The synchronization process consists of 3 phases: Code Group 
Synchronization (CGS), Initial Lane Synchronization (ILAS) and 
data transmission phase. Two special purpose signals are required
for JESD204B Subclass 1 \cite{jesd01} operation: SYSREF and SYNC. 

SYSREF is synchronous to the device clock
and it is used to synchronize an internal counter, named
Local Multi Frame Counter (LMFC), in both the
transmitters and the receivers. On the TPM board 
SYSREF and clocks are generated by a specialized AD9528 PLL and then
distributed to all the devices by means of ADCLK948 buffers.
Clock and SYSREF PCB traces to the ADCs are
length matched within 100 ps to each other, allowing the ADCs 
to sample the SYSREF on the falling edge
of the device clock with ample setup and hold margin.

In the CGS phase the FPGA issues a synchronization request
driving the SYNC signal low. When the ADCs sample SYNC low, they 
transmit a synchronization pattern consisting of /K28.5/ 
characters in the 8B10B encoding scheme. The FPGA is synchronized 
when all the transceivers receive at least four consecutive /K28.5/ 
symbols without error, at that point the FPGA drives SYNC high. 

After sampling SYNC high the ADCs wait for the next LMFC
counter wrap-to-0 event and then start the ILAS phase where
they transmit a predefined pattern. At the subsequent LMFC 
counter wrap-to-0 event the ADCs enter the Data phase
and transmit sampled data. 

On the receiving side, in the ILAS phase the FPGA 
checks for the predefined pattern and it internally delays the 
lanes in order to match the delay of the slowest lane. 
After the ILAS phase, the core starts forwarding sampled data
to the user logic at the next LMFC wrap-to-0 event.
The whole process always takes a deterministic number of 
LMFC periods.

JESD204B operation is managed in the FPGAs by the Xilinx 
JESD204 proprietary core, which handles the synchronization
autonomously and forwards ADC data towards user
logic using a simple streaming interface. Since the Xilinx
core supports up to 12 lanes, two cores sharing the same
SYSREF signal are instantiated.

\subsection{Framing and time stamping}

After synchronization the JESD204 core streams
sample aligned ADC data to the framer.
The framer packetizes the incoming data in frames that
are 864 samples long (1080~ns). At the beginning of the
observation the framer aligns the first data in the first
frame to a PPS rising edge and then it continuously
streams framed data. 

The FPGA manages two counters,
{\em sync\_time} and {\em time stamp} counters. The {\em sync\_time}
counter represents UTC time, it is set before the
observation is started and it is incremented by a PPS
rising edge. The {\em time stamp} counter is started when
the framer streams the first ADC data and it represents
the time in ns elapsed from the beginning of the observation.
At the same time, the {\em sync\_time} counter content is copied in
a local {\em start\_time} register. The combination of the 
two counters provides univocal time-stamping of each frame. 

Processed frames correspond univocally to ADC frames.
At the beginning of the observation the channelizer waits for 
a fixed amount of time, corresponding to the filter preload time, 
before outputting the first channelizer frame, and then 
produces one frame of channelized samples for each input ADC frame.
The tile beamformer processes one channelizer frame at a time. 
Therefore the time associated to each processed frame $t_f$, 
up to the tile beamformer output, can be derived by simply 
counting the frames and is not explicitly specified with 
the frame itself:

\begin{equation}
t_f = t_0 + t_1 + N_f \Delta t
\end{equation}

Here $t_0$ is the time for the start of the observation, stored in the 
{\em start\_time} register, $t_1$ is the deterministic time needed to 
preload the channelizer filter ($t_1=7560$~ns in the current implementation), 
$N_f$ is the frame number, that is reset before the beginning of 
observation, and $\Delta t=1080$~ns is the frame length. 

The station beamformer packs together groups of 256 consecutive samples,
in the packets propagated through the station beamforming chain,
and 2048 samples in the packets sent to the CSP. These packets include
a frame header, in the SPEAD format, that explicitly contains the 
observation {\em start\_time} and the relative time for the first sample 
in the packet. The header format is described in section \ref{sec:spead}.

Particular care has been taken in synchronizing with the external PPS.
As the PPS signal has an arbitrary phase with respect to the system clock,
it is sampled multiple times in a clock period, and a clock phase which
does not create ambiguities is chosen. This ensures that the internal
time reference has a fixed time delay with respect to the system time.
This has been verified by providing the same signal, locked to the system
time, to multiple boards, and checking the repetibility and stability
of the sampled signal phases across boards.

\subsection{Correction for cable mismatches}

The data stream from each antenna can be delayed by an integer number of
samples, to correct for mismatches in the length of the cables 
connecting each antenna to the TPM. The correction is almost static,
i.e. it is computed only when the cable length is explicitly measured, 
and corresponds to aligning all antennas to a source at the zenith. 
It is not changed during normal operations, or for sideral tracking.

The correction range is $\pm512$ samples, corresponding to about $\pm120$ 
meters of optical fiber, and is applied together to both polarizations 
for each antenna. The residual sub-sample delay is corrected in the 
frequency domain, in the beamforming process.

\section{Polyphase channelizer}
\label{sec:pfb}

The channelizer divides the sampled bandwidth of 400 MHz into 512
equispaced spectral channels, with a spacing of 781.25 kHz and a channel
bandwidth of 925.926 kHz. The channels overlap by a factor of
5/32 (oversampling of 32/27). The extra bandwidth allows for a channel
shape that is flat in the central region, kept in the subsequent processing,
leaving the filter transition region in the overlapping channel edges, 
discarded in the CSP. After correction for the deterministic pass-band ripple,
this produces a residual flatness in the correlated signal of less than
$\pm0.01$~dB, without discontinuities near the edges of the channels. 
To achieve the requirements for in-band flatness and out-of-band rejection,
listed in section \ref{sec:proc}, a filter order of 14336 has been used. 

The channelizer is based on the oversampling polyphase concept
\citep{harris01}, optimized for a real valued, time multiplexed signal,
and to minimize the number of FPGA resources required.  A polyphase
filter implements a Weight-Overlap-Add operation (WOLA) on the input
samples using a prototype low-pass filter, producing a vector of size
$N$. This vector  is then Fourier transformed, producing $N$ equispaced
channels ($N/2$ for a real valued signal) corresponding to the center
time of the input sample segment.  The Fast Fourier Transform (FFT)
effectively translates the prototype filter response to each channel
center frequency.  For critically sampled filterbanks, the filter and
the FFT is computed for times which are multiples of $N$ samples, but
reducing this spacing to a value $M<N$, larger output bandwidths can
be obtained.  In our case $N=1024$ and $M=864$.

\subsection{Polyphase filter}

\begin{figure}
\begin{center}
\includegraphics[width=0.9\textwidth]{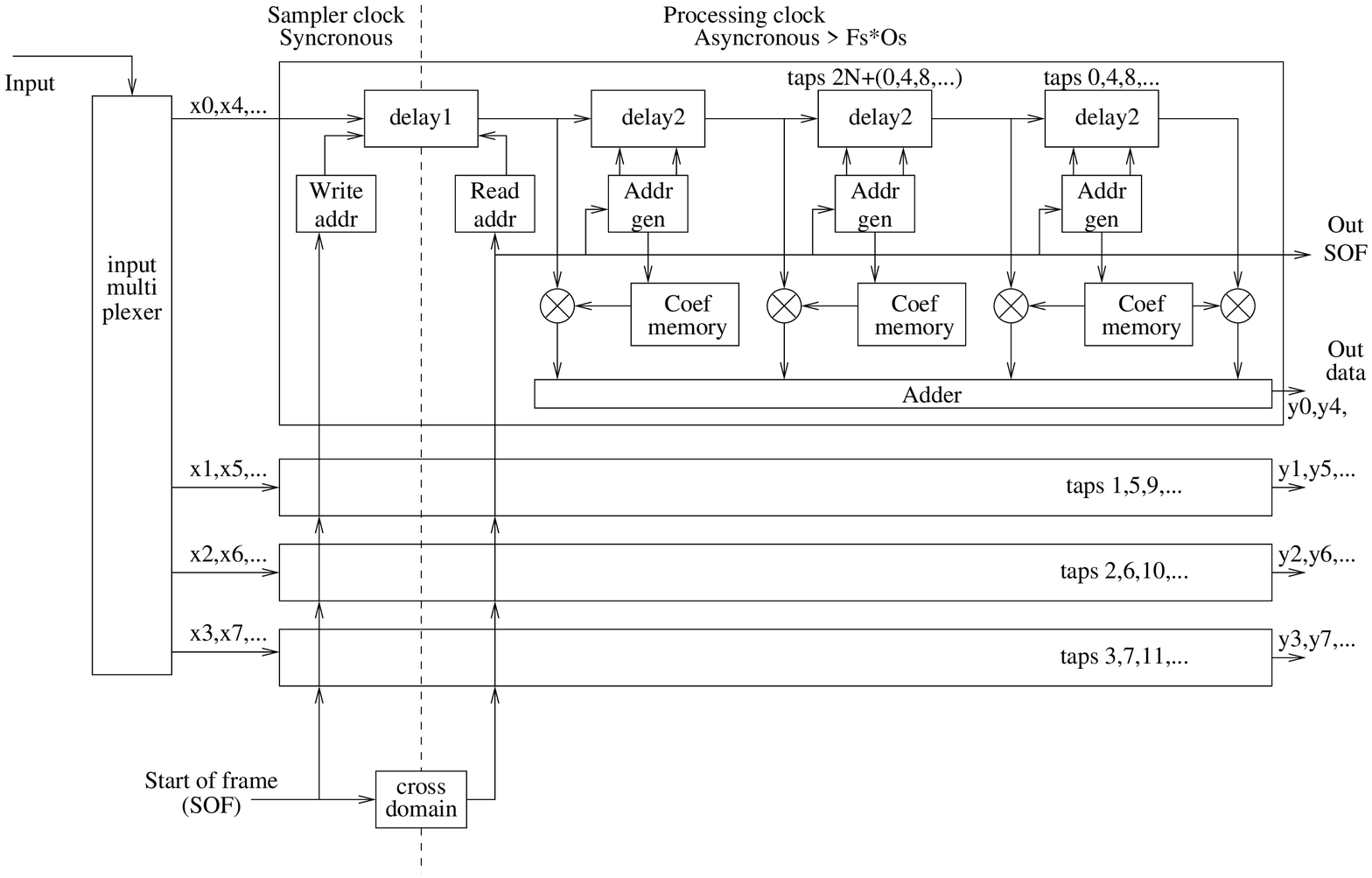}
\end{center}
\caption{
Architecture for the time multiplexed, oversampling polyphase filter
}
\label{fig:pfb}
\end{figure}

The WOLA filter architecture is shown in Figure \ref{fig:pfb}.  A first
dual port memory block produces blocks of N contiguous samples for
every frame of M input samples.  The output port runs at a higher clock
frequency, and repeats the last $(N-M)$ samples of the previous block
at the beginning of each frame.  Successive delay blocks introduce a
delay of $N$ samples, again repeating $(N-M)$ samples at the beginning
of each frame.

The structure is adapted to a time multiplexed data stream, with 4 
consecutive samples produced at each clock cycle. The whole filter 
also produces 4 consecutive values at each clock cycle. To optimize 
memory usage, a single memory 4 times wider is used for the parallel 
samples in each block. The filter tap coefficients are identical for
all the antennas, and a single memory is used for all parallel instances 
of the filterbank. As the filter is symmetric, a single dual port 
read-only memory is used to generate two symmetric sections of the filter. 
The total number of WOLA blocks is related to the filter specifications. 
In our design we used 14 blocks.

For oversampled polyphase filters, the oversampling introduces a cyclic 
phase rotation in each spectral channel that is corrected by rotating 
the filtered frame before the FFT.

\subsection{FFT block}

\begin{figure}
\begin{center}
\includegraphics[width=0.9\textwidth]{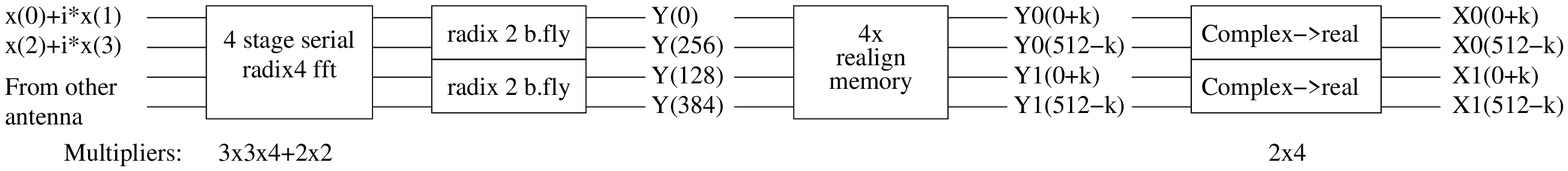}
\end{center}
\caption{
1024 point FFT core for two 4x time multiplexed real valued signals
}
\label{fig:fft4}
\end{figure}

The FFT of the real sequence is performed packing two consecutive samples
as the real and imaginary part of a single complex value, computing
a complex valued FFT of size $N/2$, and then separating channels $k$ 
and $(N/2-k)$ using the technique described in \cite{press01}. 
FPGA implementation is shown in Figure \ref{fig:fft4}.  The FFT is composed
of a serial Decimation-in-Frequency (DIF) radix-4 FFT block, that processes
simultaneously two signals (two polarizations for a single antenna),
followed by a parallel radix-2 butterfly stage and by the final 
real channels separation stage. 

The core is optimized for a signal consisting of Gaussian noise. Overflow
is possible in the last stages in the presence of monochromatic (RFI)
signals, and the affected channels are properly flagged, but does not occur
for normal radioastronomic signals or non monochromatic RFIs. The structure
is very efficient, using just 48  multipliers for two parallel FFTs of
1024 points each, sampled at 4 times the FPGA clock frequency.

We analyzed several commercial or public domain firmware cores, but the
resource usage was significantly higher. The custom core is also vendor
independent and can be easily tailored to specific SKA requirements.

\subsection{Filter design}

\begin{figure}
\begin{center}
\includegraphics[width=0.9\textwidth]{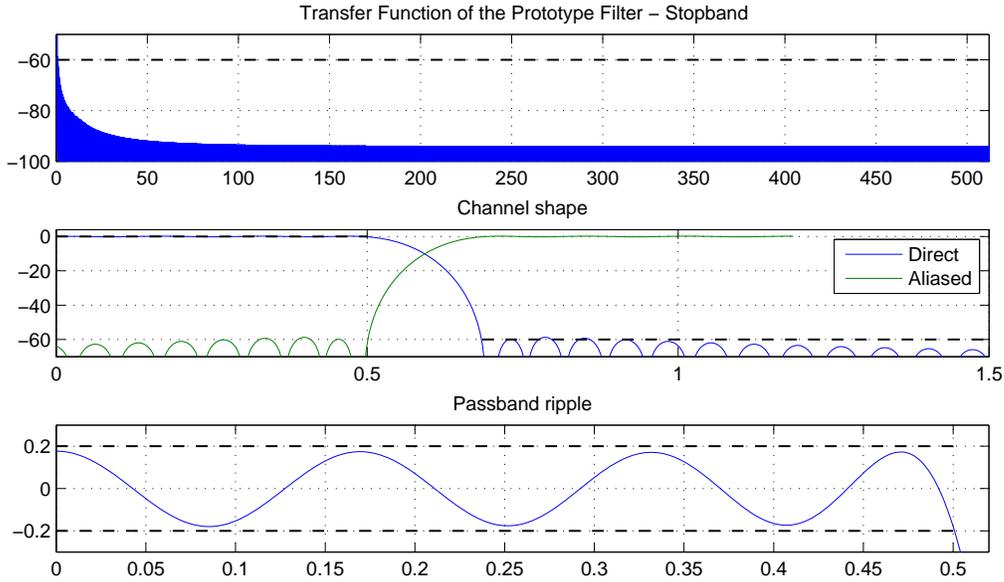}
\end{center}
\caption{
Prototype channel filter response. Frequency scale is expressed in units of 
channel-to-channel separation, vertical scale in dB
}
\label{fig:filtresp}
\end{figure}

The filter block, being repeated for each signal, is the most resource
intensive structure in the design. Therefore filter design has
required particular care, in order to minimize filter size for given
specifications.  Common design techniques for filters of this size
usually involve windowing of an ideal passband filter. This produce
very high stop-band rejection, above what is effectively needed, at
the expense of a larger size. Equiripple design techniques, like the
Remez-mcClellan algorithm \citep{parks01}, usually do not converge
for filters of order beyond a few hundreds. We adopted the technique
of designing a relatively short (1/16 of the final size) equiripple
filter using the Remez-mcClellan algorithm, and interpolating the result
using a FFT based interpolator \citep{comore01}.  The discontinuity
present at the edges of the equiripple filter has been deleted before
interpolation, and reintroduced in the interpolated result. We used a
custom C implementation of the algorithm, which has been proved more
robust than the one in the Matlab filter package.

The resulting filter shape is shown in Figure \ref{fig:filtresp}. The passband
ripple has been set to $\pm0.17$~dB, the stop-band attenuation is greater
than 60 dB at the beginning of the aliased region and drops to 86 dB
for most of the stop-band. A total of $14N$ (14336) filter taps are required for
a polyphase filter with the required oversampling factor of 1.185. For
comparison, a filter with the same performances designed with conventional
techniques has a size of at least $18N$.


\section{Beamformer and calibration}
\label{sec:bmf}

Station beamforming is performed in the frequency domain, by phase
referencing each antenna to a common station phase center and summing
together the 256 antennas composing the station. Receiver, atmospheric and
polarization calibrations are also applied to each signal before summing
them together.  The sum is performed hierarchically: the 8 stations in
each FPGA are summed together, then odd and even channels for the whole
tile are exchanged among FPGAs and summed in a tile partial beam, 
and finally each partial beam is summed to a traveling packet that 
runs across the tiles composing the station.

\subsection{Tile beamformer and calibration}

The tile beamformer is described in more detail in \cite{comore02}.
SKA-low specifications require the beamformer to produce up to up
to 8 independent beams, placed anywhere in the sky, and up to 16
independently tunable frequency regions (sub-bands), placed anywhere
in the digitized band, for a total bandwidth of 300 MHz (384 channels). 
Each sub-band can be assigned to any beam, and can repeat the same frequency region
(e.g. 8 identical sub-bands for 8 beams at the same frequency).   
Due to hardware limitations, the sub-band
width is constrained to a multiple of 8 spectral channels (6.25 MHz),
and its position can be set with a granularity of 2 spectral channels
(1.5625 MHz).

The selected channels are then multiplied by a phase proportional
to the sample frequency and antenna geometric delay. The delay is
dynamically computed using an initial delay value and a delay rate.

Before beamforming, the signals from each antenna are corrected for 
amplitude and phase instrumental response, for atmospheric propagation 
effects, and for beamforming tapering. The correction is 
expressed by a complex polarization matrix for each frequency channel,
antenna and beam. Signals are also equalized in 
amplitude, to allow for requantization with 8+8 bits/sample without 
significant quantization losses.

Thus the beamformed signals $S(t,\nu,b)$ for one station, two polarizations,
for the frequency channel $\nu$ and beam $b$ is given by: 

\begin{equation}
\begin{array}{r c l}
\left( \begin{array}{c} S_h(t,\nu, b) \\ S_v(t,\nu, b) \end{array} \right) 
&=&
\sum_a \exp\left(2\pi j\, \nu\, \tau(t,a,b) \right)
\left( \begin{array}{cc} 
  C_{hh}(\nu,a,b) & \,\,C_{hv}(\nu,a,b)\\
  C_{vh}(\nu,a,b) & \,\,C_{vv}(\nu,a,b) \end{array} \right)
\left( \begin{array}{c} A_h(t,\nu, a) \\ A_v(t,\nu, a) \end{array} \right)\\
\strut& &\\
\tau(t,a,b)&=&\tau_0(a,b)+t\, \dot\tau (a,b)
\end{array}
\label{eq:bmf}
\end{equation}

where $C(t,\nu,b)$ is the beamformed signal for beam $b$, 
$A(t,\nu,a)$ is the channelized signal for antenna $a$, 
$\tau(t,a,b)$ is a linear approximation for the geometric delay of 
antenna $a$ relative to beam direction $b$, and $C(\nu,a,b)$ is a 
correction matrix. Subscripts $h,v$ refer to the two linear polarizations. 
All quantities, except the delay, are complex, 
and specified for the 384 selected combinations of frequency and beam
directions. The corrections are referred to the center of each frequency channel
and decimated sample interval. The phase error increases 
towards the channel edges. For a channel width of 781.25~kHz, 
an uniform circular station with a diameter of 35 meters, and pointing 
up to 45 degrees from the zenith, the maximum phase error
is 0.1 radians and the corresponding beam decorrelation for a uniform
circular station is 0.25\%, or $-0.01$~dB. This has been test using a 
simulated linear array, with results in figure \ref{fig:beamform_loss}.

The correction matrix $C$ is composed of several terms: 
\begin{itemlist}
\item the correction for the (complex) antenna/receiver frequency 
response, including the subsample cable length mismatch
\item the inverse of the Jones matrix relating the polarization in 
the sky to the measured fields at the two antenna polarization ports,
including the parallactic rotation
\item the desired antenna tapering amplitude for beamforming
\item an equalization term, common to all antennas and polarizations in 
the station, to provide the CSP with a signal of constant amplitude. 
\end{itemlist}

All quantities are calculated externally to the TPM, and 
expressed as integer values. Linear delay is 
updated every 1024 channelized samples (1105.92~$\mu$s). Phase is 
expressed with a resolution of 4096 steps/turn, delay with a resolution 
of 153~fs in a range of $\pm80$~ns ($\pm24$~m), and delay rate 
with a resolution 
of 8.4~fs/s and a range of $\pm17$~ns/s. For a source moving at a sidereal 
rate, delay and delay rates can be updated every 2 minutes before the 
phase error becomes significant. 

Matrix $C$ is expressed as a complex mantissa, with $16+16$ bits of accuracy, 
and a 3 bit exponent specified every 8 frequency channels for each antenna. 
The mantissa is updated at the calibration cycle interval, currently specified 
as 10 minutes. 

Operations in equation \eqref{eq:bmf} is performed in the following order: 
\begin{itemlist}
\item Channelized samples (18 bit values) are scaled by the exponent 
of $C(\nu)$ and requantized to 12 bit accuracy;
\item Samples are multiplied by the complex exponential for the 
geometric delay;
\item Samples are multiplied by the mantissa of the matrix $C(\nu)$, 
and requantized to 8 bit accuracy;
\item The sum for beamforming is performed using 16 bits, and the 
16 bit result is rescaled and requantized to 8 bit accuracy for the CSP. 
\end{itemlist}

\subsection{Station beamformer}

The tile beamformer produces partial sums of 16 antennas, organized in frames of
one time sample and 150 MHz of frequency channels. 
These partial beams are stored in local memory, in blocks of up to 
about 0.23 seconds, and retrieved as contiguous frames of 128 time samples
for 8 channels and 2 polarizations. Each sample is represented as a $16+16$
bit complex value. 

The TPMs composing a station are logically organized in a network
chain, with each TPM sending frames for the partial beam to the
next one in the chain. By changing the address of the next TPM, 
it is possible to define the tiles composing each station. 
The first TPM in the chain
retrieves sequentially the frames for the same channels and for the 
whole time block, then send them to the next one. Every other TPM, 
upon receiving
a frame, retrieves from memory the corresponding frame and adds it to the 
traveling partial sum. The last TPM in the station chain accumulates 
16 frames, forming 8 frames with 2048 samples and a single frequency 
channel, and sends each frame to a separate section of the CSP. Samples
sent to the CSP are represented as $8+8$ bit complex values. 
In this way each section of the CSP receives from each station a 
contiguous stream of samples for a single channel at a time. 

Each frame, in both formats, has an associated header in the format 
specified in section \ref{sec:spead}.


\section{SPEAD formatter}
\label{sec:spead}

Beamformer and channelizer data is transferred between TPMs using
SPEAD \cite{manley01}, an application-layer protocol developed for the
CASPER (Collaboration for Astronomy Signal Processing and Electronics
Research)\footnote{https://casper.berkeley.edu/} and widely adopted by
the radio astronomy community.  Data is streamed in and out of the SPEAD
formatter using the AXI4-Streaming Protocol \cite{arm01}. In addition
AXI4-Stream $t_{USER}$ side channels and additional ports are used to
populate and construct the SPEAD header and the total length of the
data payload is configured via AXI4-Lite Memory-Mapped \cite{arm01}
registers and is expected to be a $2^{N}$ value. Incoming $t_{DATA}$ is
expected to be equal in size to, or a multiple in size of the $2^{N}$
value in the configuration register. Incoming SPEAD payload data is
valid when $t_{VALID}$ is asserted and the final value is determined
when $t_{VALID}$ and $t_{LAST}$ are both asserted. There is no internal
error checking of the size of incoming $t_{DATA}$ streams.

The SPEAD formatter uses two First In First Out (FIFO) memories on the
slave side of the AXI4-Stream Interface. The first FIFO is dedicated to
the payload data and has two functions; the first function is to
allow the crossing of clock domains and to handle changes in incoming
$t_{DATA}$ data width to match the clock/data width of the up-stream
component(s) generating the data; second function is to act as a shallow
buffer for the incoming payload data while the header fields
are being assembled. The second FIFO is used to hold the header
fields as they are constructed. Once the header is complete an
internal controller State-Machine places the header followed by the
payload data into a third FIFO to present the combined, complete
SPEAD packet to the output AXI4-Stream master interface. This final
FIFO allows the crossing of clock domains and to handle width changes
between the internal $t_{DATA}$ width of 64 bits, defined by the SPEAD
Specification \cite{manley01}, and to match the clock/data width of the
down-stream component(s) connected to the SPEAD formatter.

Using additional firmware components data can be transferred FPGA-to-FPGA
or FPGA-to-Network Interface, either via a network switch or directly
to a server's network interface. This is achieved by encapsulating the
SPEAD application-layer stream data into single or multiple User Datagram
Protocol (UDP) packets, where each packet contains its own SPEAD header;
in this example the maximum data payload stream is limited to
8192 bytes plus the header, an additional 72 bytes. This due to
the custom 10Gb Ethernet UDP Media Access Control (MAC) firmware used
on the TPM FPGAs which is not discussed in this paper.

The formatter firmware is capable of implementing SPEAD Version 4,
SPEAD-64-40 and SPEAD-64-48 implementation of SPEAD set at compilation
time via Very-high-speed-integrated-circuits Hardware Description Language
(VHDL) generics and is capable of generating four header types which are
set at compilation via VHDL generics depending on the source of the data
stream to be transmitted and its required destination. To fully utilize
the customized {\it immediate}\footnote{In the context of SPEAD the terms
{\it immediate} and {\it absolute} refer to using the Most Significant
bit (MSb) of the 64 bit wide header field as an indicator to flag
whether the item field contains the data defined by the item ID ({\it
immediate}, $MSb = 1$) or the offset address of the location within the
payload data ({\it absolute}, $MSb = 0$).} item fields SPEAD-64-48
is used in all implementations of the formatter used on the TPMs.

\subsection{SPEAD header}

In order to differentiate between the different types of streaming data
generated by the TPMs a unique SPEAD header of 72 bytes is generated
and transmitted with the data.  The header format loosely follows the
recommended practices \cite{manley02}.  The required Heap offset (item
ID 0x0003), \cite{manley01}, has been omitted and some item IDs are
recommended to use {\it absolute} address references to the corresponding
location in the payload data.  Instead the {\it immediate} method
has been used in order to reduce the header size and its effect
on the overall bandwidth of the system.

Regardless of source of the incoming data-stream, the first five header fields
following the header identifier field are common
in all SPEAD formatter implementations and deviation from the SPEAD recommended 
practices are shown in Table \ref{spead:tbl1}.\footnote{If the SPEAD recommended 
practices were followed the SPEAD header would have a total size of 88 bytes, 
plus the addition of a further 16 bytes of corresponding data embedded into the 
payload data, a total overhead of 104 bytes/packet regardless 
of the size of the payload data. Hence the non-standard optimizations made.}

\begin{wstable}[h]
\caption{First five SPEAD formatter header fields}
\begin{tabular}{@{}cccc@{}} \toprule{a}
Field Index & Item ID & Hex Code & Item Field(s) \\ \colrule
1 & Heap Counter & 0x0001 & Logical Channel ID\tnote{a}, Packet Counter\tnote{b} \\
2 & Packet Length & 0x0004 & Packet Payload Length\tnote{c} \\
3 & Reference Time & 0x1027 & Unix Format Reference Time (s)\tnote{d} \\
4 & Timestamp & 0x1600 & Timestamp (ns)\tnote{e} \\
5 & Center Frequency & 0x1011 & Frequency (Hz)\tnote{f} \\ 
\botrule
\end{tabular}
\begin{tablenotes}
\item[a]2 bytes
\item[b]4 bytes
\item[c]6 bytes
\item[d] Unix Time is 4 bytes but resized to 6 bytes to fill this field
\item[e]No corresponding Time-Stamp scale (item ID 0x1027) used for Time-Stamp 
value as defined in \cite{manley02}. Always assumed to be nano-second.
\item[f]Immediate, integer value of Center Frequency - not an absolute addressed
IEEE float 64 as defined in \cite{manley02}
\end{tablenotes}
\label{spead:tbl1}
\end{wstable}

Upon receiving the first $t_{VALID}$ signal the formatter registers each of
the header field/sub-field values required to complete the header. The source of
these field values differ and can originate from: 
\begin{romanlist}[iii]
\item VHDL generics configured at compile time; 
\item AXI-4 Stream side-channels ($t_{USER}$, 4 bytes wide);
\item dedicated inputs to the SPEAD formatter VHDL component. 
\end{romanlist}

The sources for each field/sub-field are shown in Table \ref{spead:tbl3}.
In order to simplify the memory addressing in software, all header input
sources, regardless of source, are resized by the SPEAD formatter to
fit the number of bytes allocated in the header item fields.

In addition to using $t_{USER}$, $t_{ID}$ (1 byte wide)
is used to connect to down-stream UDP MAC components and is used as an
index to a preconfigured, via the AXI4-Lite Memory Mapped interface, Look Up Table (LUT) 
containing a list of MAC Addresses; Internet Protocol (IP) Addresses; and UDP Destination Ports.
This allows TPMs to target specific destinations through network switches to:
\begin{romanlist}[iii]
\item create the beamforming ring;
\item send data to the CSP or the calibration subsysrem of the 
Local Monitor and Control system; 
\item connect to station correlation hardware.
\end{romanlist}

\begin{wstable}[h]
\caption{SPEAD header field sources}
\begin{tabular}{@{}ccc@{}} \toprule
Field & Sub-Field & Source \\ \colrule
SPEAD Header & Magic Number\tnote{a} & VHDL Generic \\
SPEAD Header & SPEAD Version\tnote{a} & VHDL Generic \\
SPEAD Header & Item Identifier Width\tnote{a} & VHDL Generic \\
SPEAD Header & Heap Address Width\tnote{a} & VHDL Generic \\
SPEAD Header & {\it Reserved}\tnote{b} & VHDL Generic \\
SPEAD Header & No. of Header Items\tnote{a} & VHDL Generic\\
Heap Counter & Logical Frequency Channel ID\tnote{b} & $t_{USER}$[28..13] \\
Heap Counter & Packet Counter\tnote{c} & Dedicated Input \\
Packet Length & Packet Length\tnote{d} & Dedicated Input \\
Reference Time & Reference Time\tnote{d} & Dedicated Input \\
Timestamp & Timestamp\tnote{d} & Dedicated Input \\
Center Frequency & Center Frequency\tnote{d} & Calculated using Physical Frequency Channel ID ($t_{USER}$[8..0])\\
CSP Channel Info & Beam ID\tnote{b} & $t_{USER}$[12..9] \\
CSP Channel Info & Physical Frequency ID\tnote{b} & $t_{USER}$[8..0] \\ 
CSP Antenna Info & Sub-Array ID\tnote{a} & Dedicated Input \\
CSP Antenna Info & Station ID\tnote{b} & Dedicated Input \\
CSP Antenna Info & No. of Contributing Antenna\tnote{b} & Dedicated Input \\
\botrule
\end{tabular}
\begin{tablenotes}
\item[a]1 byte
\item[b]2 bytes
\item[c]4 bytes
\item[d]6 bytes
\end{tablenotes}
\label{spead:tbl3}
\end{wstable}

The Heap counter field (item ID 0x0001) is constructed using the logical
channel ID, 2 bytes wide, and an externally incremented 4 byte wide
packet counter. The logical channel ID is the identifier of the channel
in one beam, which may contain a range of physical channels. The packet
counter should be incremented for each packet being generated by
the SPEAD formatter.  This gives each packet a unique identity
and can be used by the receiving logic on network connected systems,
either FPGAs or server hardware, to handle out-of-order receiving of
data and to determine the loss of packets.

The Center Frequency field (item ID 0x1011) is calculated by the SPEAD
formatter by multiplying the channel spacing of 781.25 kHz, described in
section \ref{sec:pfb}, with the Physical Frequency channel ID taken from
$t_{USER}[8..0]$. The calculated value is placed in the Center Frequency
field as an integer value.




\subsubsection{Additional SPEAD formatter header fields}
In order to fulfill complete TPM functionality there are currently seven header 
field IDs which are used for transferring data: 
\begin{romanlist}[iv]
\item between TPMs; 
\item to CSP; 
\item to LMC; 
\item to local station correlation hardware. 
\end{romanlist}

The SPEAD header item IDs, and their corresponding positions in the 
header, denoted by the Field Index, are shown in Table \ref{spead:tbl2}.

\begin{wstable}[h]
\caption{Additional SPEAD formatter header fields}
\begin{tabular}{@{}cccc@{}} \toprule
Field Index & Item ID & Hex Code & Item Field(s) \\ \colrule
6 & LMC Raw Data Info & 0x2000 & {\it Reserved}\tnote{c}, Antenna Start ID\tnote{a}, No. of Included Antenna\tnote{a} \\
7 & LMC TPM Info & 0x2001 & TPM ID\tnote{b}, Antenna Station ID\tnote{b}, {\it Reserved}\tnote{b} \\
6 & LMC Channel Info & 0x2002 & Start Channel ID\tnote{c}, No. of Included Channels\tnote{a}, Antenna Start ID\tnote{a} \\
7 & LMC Antenna Info & 0x2003 & TPM ID\tnote{b}, Antenna Station ID\tnote{b}, No. of Contributing Antenna\tnote{b} \\
6 & CSP Channel Info & 0x3000 & {\it Reserved}\tnote{b}, Beam ID\tnote{b}, Physical Frequency ID\tnote{b} \\
7 & CSP Antenna Info & 0x3001 & {\it Reserved}\tnote{a}, Sub-Array ID\tnote{a}, Station ID\tnote{b}, No. of Contributing Antenna\tnote{b} \\
8 & CSP Sample Vector & 0x3300 & Payload Offset\tnote{d} \\
\botrule
\end{tabular}
\begin{tablenotes}
\item[a]1 byte
\item[b]2 bytes
\item[c]4 bytes
\item[d]6 bytes - Fixed Value 0x000000
\end{tablenotes}
\label{spead:tbl2}
\end{wstable}

\subsubsection{SPEAD formatter payload}
For CSP sample vectors (item ID 0x3300), the payload data is presented in
8 byte words to the formatter. Starting from the Least Significant
Byte (LSB), 1 byte time samples for each of the two polarities and
the real and complex parts are transferred in the order shown in Table
\ref{spead:tbl4}.

\begin{wstable}[h]
\caption{SPEAD formatter payload fields for CSP Sample Vector (item ID 0x3300)}
\begin{tabular}{@{}ccccccccc@{}} \toprule
Word & MSB & & & & & & & LSB \\
Index & Byte 7 & Byte 6 & Byte 5 & Byte 4 & Byte 3 & Byte 2 & Byte 1 & Byte 0 \\ \colrule
0 & $V_{POL} \Re_{[1]}$ & $V_{POL} \Im_{[1]}$ & $H_{POL} \Re_{[1]}$ & $H_{POL} \Im_{[1]}$ & $V_{POL} \Re_{[0]}$ & $V_{POL} \Im_{[0]}$ & $H_{POL} \Re_{[0]}$ & $H_{POL} \Im_{[0]}$ \\
1 & $V_{POL} \Re_{[3]}$ & $V_{POL} \Im_{[3]}$ & $H_{POL} \Re_{[3]}$ & $H_{POL} \Im_{[3]}$ & $V_{POL} \Re_{[2]}$ & $V_{POL} \Im_{[2]}$ & $H_{POL} \Re_{[2]}$ & $H_{POL} \Im_{[2]}$ \\
2 & $V_{POL} \Re_{[5]}$ & $V_{POL} \Im_{[5]}$ & $H_{POL} \Re_{[5]}$ & $H_{POL} \Im_{[5]}$ & $V_{POL} \Re_{[4]}$ & $V_{POL} \Im_{[4]}$ & $H_{POL} \Re_{[4]}$ & $H_{POL} \Im_{[4]}$ \\
... & ... & ... & ... & ... & ... & ... & ... & ... \\
1023 & $V_{POL} \Re_{[2047]}$ & $V_{POL} \Im_{[2047]}$ & $H_{POL} \Re_{[2047]}$ & $H_{POL} \Im_{[2047]}$ & $V_{POL} \Re_{[2046]}$ & $V_{POL} \Im_{[2046]}$ & $H_{POL} \Re_{[2046]}$ & $H_{POL} \Im_{[2046]}$ \\
\botrule
\end{tabular}
\label{spead:tbl4}
\end{wstable}
A total of 2048 time samples, representing the real and imaginary parts of 
a complex number for two polarizations, are streamed per SPEAD packet.

\subsection{SPEAD receiver}
To accommodate FPGA-to-FPGA connectivity a SPEAD receiver component has been
developed. There is only one version of this component and it is able to 
determine the source of the incoming data-stream by examining the SPEAD 
header fields before outputting the payload data on the AXI4-Streaming 
master interface. SPEAD header field items are placed into the equivalent 
source locations defined by the SPEAD formatter. That is, the AXI4-Stream 
master $t_{USER}$ is regenerated and the equivalent dedicated outputs are used. These 
master side outputs for header elements are registered on the
first $t_{VALID}$ assertion and held for the duration of the expected
streaming data payload length, determined from the corresponding 
header field, which is delimited by the final $t_{VALID}$ and $t_{LAST}$
assertion.  FIFOs are used on the AXI4-Streaming slave and master side
interfaces to allow the crossing of clock domains and to handle changes
in the internal $t_{DATA}$ width of 64 bits, again defined by the SPEAD
Specification \cite{manley01}, to make the output width wider or narrower
to match the clock/data width of the up-stream/down-stream component(s)
connected to the SPEAD receiver.

\section{Diagnostic and calibration functions}
\label{sec:diag}

The TPM provides some diagnostic and calibration features. 
These features are used to set the appropriate gain in various points across
the signal path, and to aid the station calibration algorithm. 

More sophisticated features, including a test pattern generator, 
long frame ADC data capture, or a simple correlator, are present in 
dedicated personalities, that can be loaded in the FPGAs in maintenance mode. 

\subsection{ADC total power}

The signal sampled from the ADC must remain in a definite range, 
with a Root Mean Square (RMS) amplitude comprised from 15 to 30 
ADC units, in order to guarantee proper operations. 
Lower amplitude levels 
result in excess quantization noise at the high end of the spectrum, 
while higher amplitudes introduce nonlinearities due to clipping 
of the noise-like signal. Considering the possibilities of a sudden 
variation of the input signal due to RFI, it is advisable to keep 
the signal level around 19 ADC units. 
Therefore an almost continuous monitoring of the digitized total power
is necessary. 

Each ADC input has a dedicated total power detector, that integrates
the digital power over a predefined number of input frames. The
results are read by the monitor and control subsystem. The 
integration time can be configured from about 1~ms to a few seconds.
 
\subsection{Coarse spectrum}

The channelizer output form each antenna can be used to compute a coarse
resolution spectrum, with a resolution of one spectral channel (781.25 kHz). 
At each time it is possible to compute the autocorrelation (total power) spectrum of 
one arbitrary antenna, or the cross spectrum between two arbitrary antennas. 
The coarse spectrum is used as an aid in the calibration procedure, 
while the cross correlation is used for diagnostic purposes. 

\subsection{Channelized data capture}

Station level calibration is performed by cross correlating channelized 
samples from all antennas, and comparing the results with the expected 
results from a sky model. Details of the proposed calibration techniques 
are presented elsewhere, and the final calibration algorithm is still under study,
but all these techniques rely on samples captured 
after the coarse channelization. Calibration is performed on a single 
spectral channel at a time, over a representative subset of the total 
observed channels. Calibration curves for each antenna are then derived
by interpolating the calibration solution for the observed channels. 

The channelized data capture module extracts samples for all antennas 
and polarizations for a single channel, and frames them in a SPEAD 
block for 128 8+8 bit complex samples, 32 signals. 
The resulting packets can be sent both on 
the control interface or on the high speed optical link. 
The average data rate is 0.5 Gbps, i.e. about 50\% of the bandwidth 
of the control interconnection, or 1.3\% of the bandwidth of the high 
speed link. 

Captured data can be received using a fast UDP data capture 
library. Data format has been initially checked using the program {\em 
wireshark},
to confirm that the header format, data payload and size was conform to
the expectations.

\section{Firmware testing, implementation and performance}
\label{sec:results}

The firmware is composed of individual library modules, Each module
is tested separately using a VHDL simulation engine with a specific
testbench. The testbench generates a stimulus signal, that is either
a tone, a set of tones, or a pseudo-random white noise. The control
interface is simulated using a VHDL Axi4lite master emulator, that reads
and writes data to disk files, Both input and output signals are recorded
on files. A Matlab code then reads the input signals, process them in a
functional model and compares the results to the VHDL simulation outputs.
We use both a "golden standard" model, with floating point accuracy, and 
a fixed point model with requantization of the signal 
in a few critical points. The model is 
not exact at the bit level, but differences are limited to details in the 
actual rounding. Noise performance is then analyzed statistically, comparing 
model and simulations with different quantization stages turned on and off. 

\begin{figure}
\begin{center}
\includegraphics[width=0.8\textwidth]{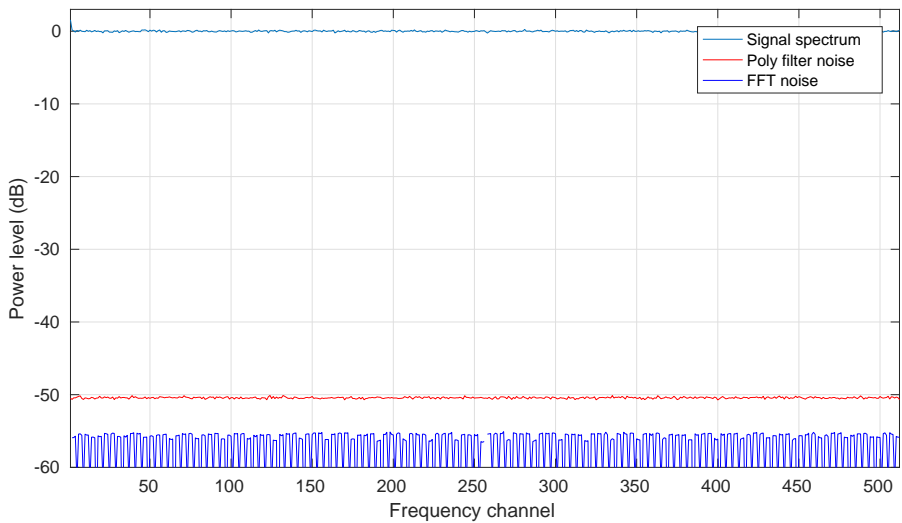}
\end{center}
\caption{VHDL simulation results for the polyphase channelizer. From top:
channelized power, quantization noise power due to requantization 
in the polyphase filter and in the FFT. Power scale in dB with respect 
to the average power}
\label{fig:model}
\end{figure}

An example of a simulation result, for the polyphase filterbank, is shown in 
figure \ref{fig:model}. A pseudo-random white Gaussian noise, with a RMS 
level of 26 ADC units,  has been used as the input signal.
The measured power level in each spectral channel is shown in the top
curve, both for the VHDL simulation and for the Matlab model (they are 
nearly identical). The quantization noise due to the  
polyphase filter and to the FFT stage has been derived by comparing 
the simulation and the model, and the noise power for these has been 
plotted respectively in the remaining two curves,
about -51 dB and -55 dB with respect to the signal level.

\subsection{Implementation results}

A complete model for the firmware has also been generated, using a 
socket VHDL interface to control it form the actual control software. 

Once all the modules have been validated at the functional level, 
the firmware has been implemented in a Xilinx XCKU040. 
Most of the synthesis has been performed using the Xilinx Vivado tool, 
but for some critical module an external synthesis tool (Synplify) 
has been used. 

Resource utilization is summarized in table \ref{tab:resources}, and refers for 
8 antennas (16 ADC channels), beamforming, and high speed Ethernet interface. 

\begin{table}[htb]
\begin{center}
\begin{tabular}{| l | c | c | c | c |}
\hline
Resource    & LUTs & Registers & Block RAM & DSP48 \\
\hline
Channelizer & 54k & 91k        & 279       & 1343 \\
Tile beamformer & 21k & 45k & 97 & 400 \\
Station beamformer & 2k & 2k & 62 & 5 \\
Other & 73k & 100k & 97 & 0 \\
\hline
Total & 150k & 238k & 535 & 1748 \\
\smallskip\% of available  & 62\% & 49\% & 89\% & 91\% \\
\hline
\end{tabular}
\end{center}
\caption{Resource utilization for one FPGA}
\label{tab:resources}
\end{table} 

The channelizer uses 81.5 multipliers per signal (163 every two signals), 
56 for the polyphase filter, and 25.5 for the FFT.
The synthesis tool has used some 
extra resources in the arithmetic blocks to improve the FFT closure time, 
for a total of 28 DSP blocks per FFT. The polyphase filter uses 151 RAM blocks, 
and the FFT 128, 8 (16 Kbyte total) per signal. 

The beamformer uses 50 multipliers per signal. 2 are used to compute the
current phase adjustment, 32 (8 complex) for the Jones matrix calibration
and 16 (4 complex) for the phase adjustment.

Although individual components could be clocked at higher speed,
e.g. the channelizer works for sample frequency up to 240 MHz,
the whole design runs just above the required 200 MHz input frequency
(800 MHz ADC sampling rate), A better placement is
under study to increase the maximum clock speed.

\subsection{Channelizer performance}

The TPM has not yet been used in the field, and all tests have been performed
using simulated signals in the laboratory, either an internal digital sinewave
generator or external analog signals. 

\begin{figure}
\begin{center}
\includegraphics[width=0.9\textwidth]{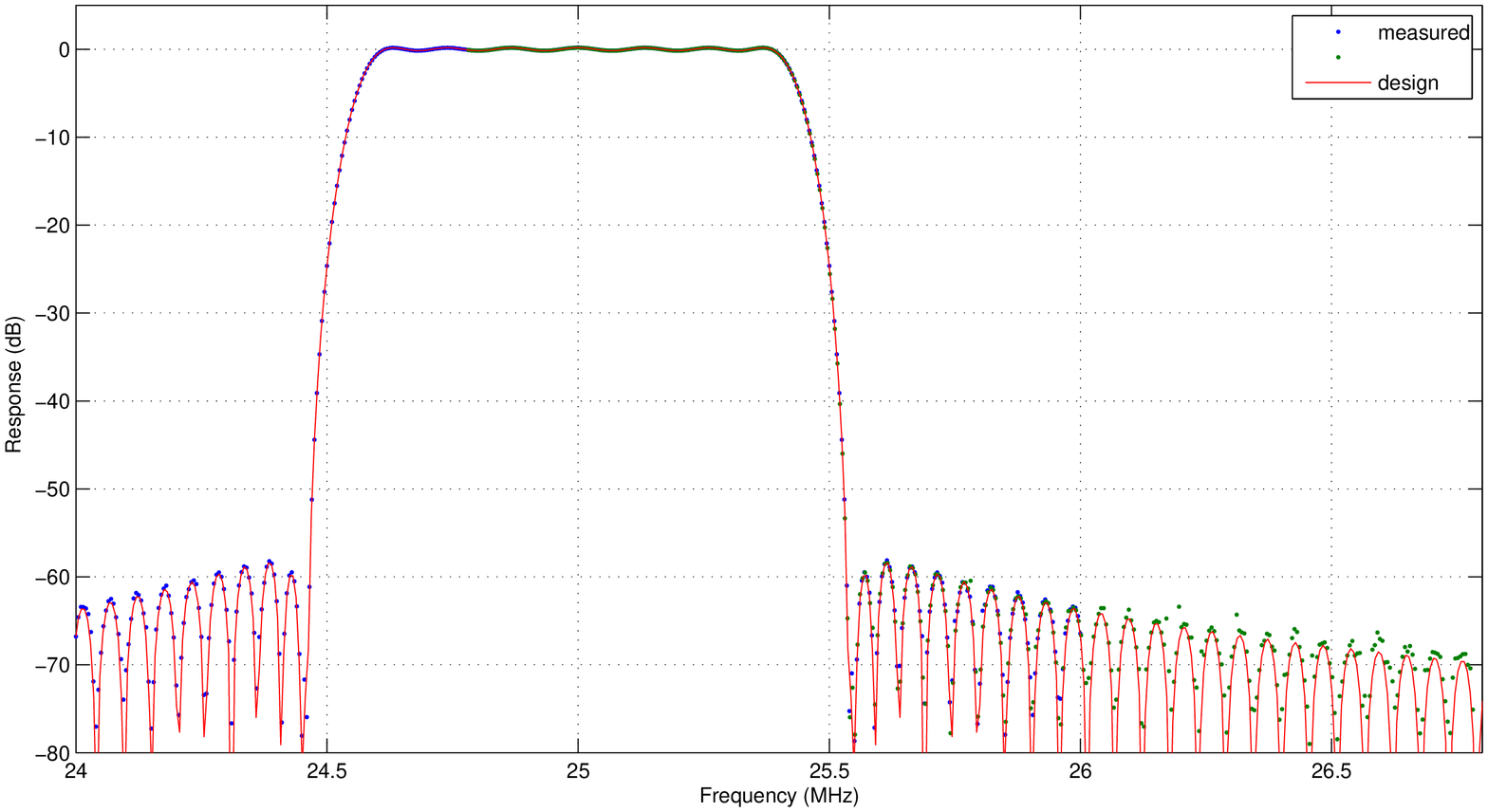}
\includegraphics[width=0.9\textwidth]{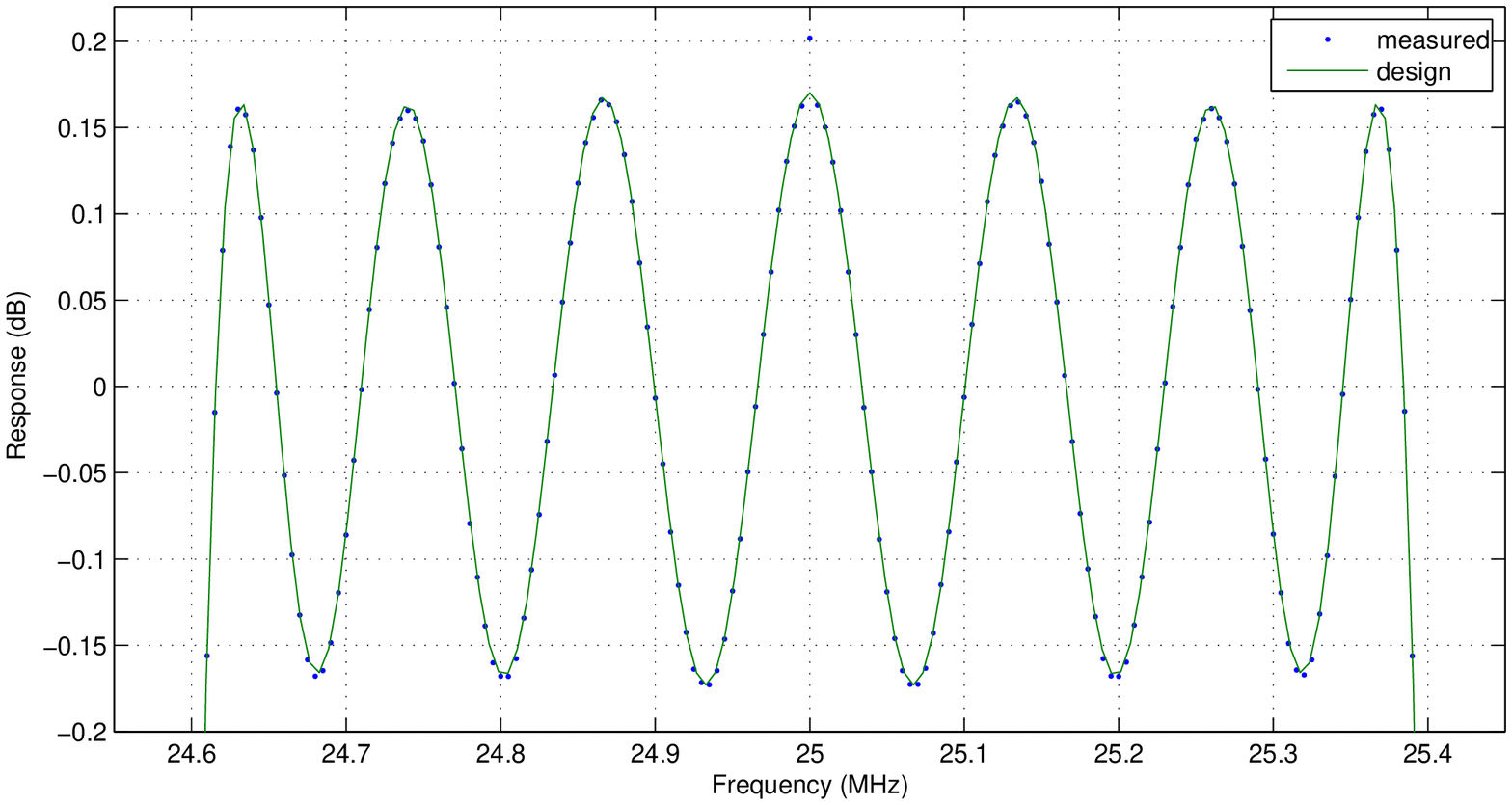}
\end{center}
\caption{
Measured channel filter response. Lower figure 
is a zoom over the passband region. The continuous line represents 
the computed filter response, the points the measured values. 
Vertical scale in dB}
\label{fig:response}
\end{figure}

The channelizer filter shape have been analyzed using the internal 
sinewave generator, that produces a sinewave at full 8 bit scale
with an arbitrary frequency.  
The signal frequency has been swept across three output channels, and the
results compared with the filter bandpass of figure \ref{fig:filtresp}. The
results are visible in figure \ref{fig:response}. 

The correspondence is very good, up to -75 dB. Below this level the quantization
noise of the 8 bit digitized sinusoid produces a uniform spectral background
noise, and a different signal must be used to assess larger filter
attenuations.  

\begin{figure}
\begin{center}
\includegraphics[width=0.9\textwidth]{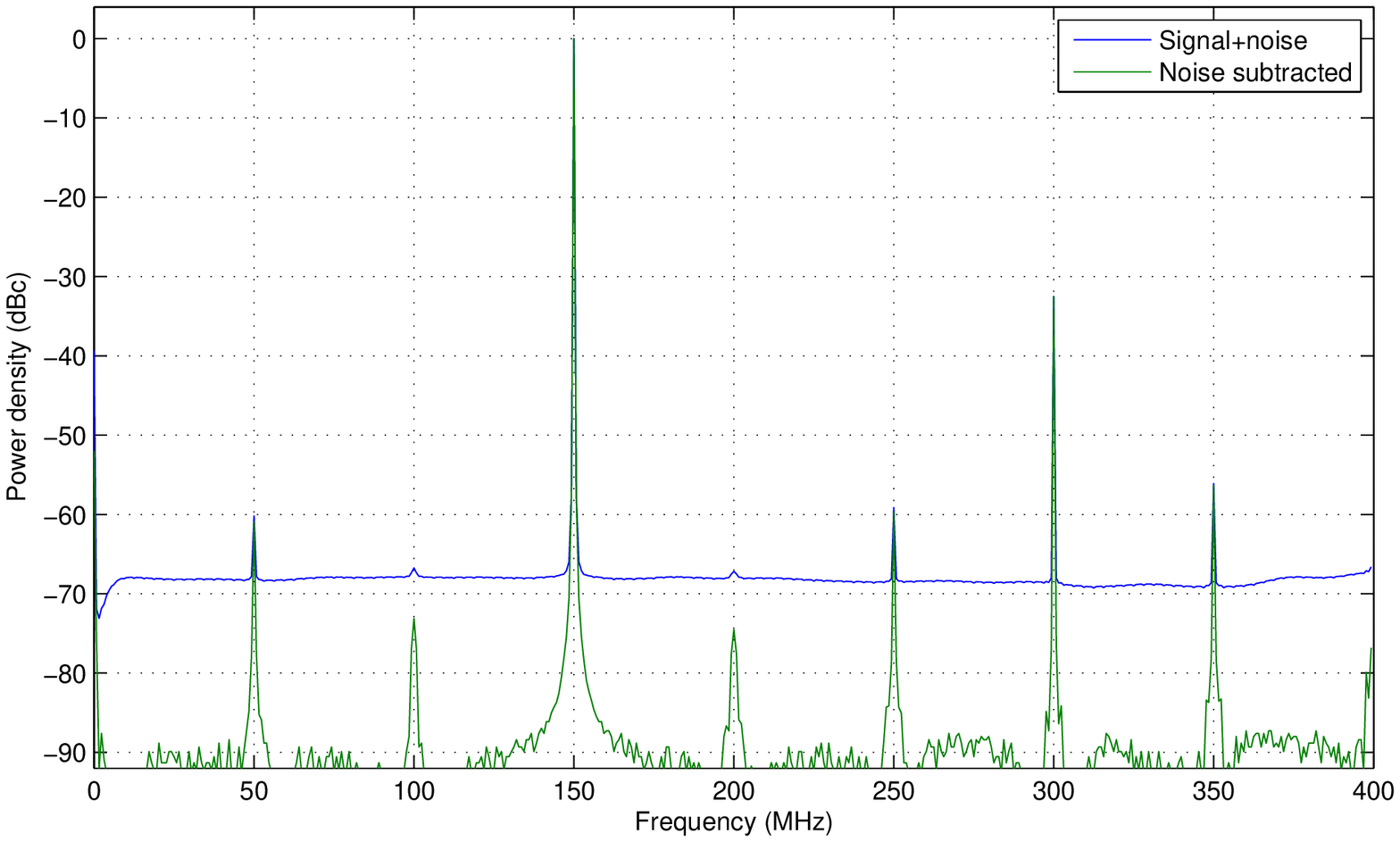}
\end{center}
\caption{
Channelized power spectrum of an analog tone signal plus noise, with 
and without subtraction of the noise power}
\label{fig:response2}
\end{figure}

A monochromatic tone from a good quality signal generator, 
with a frequency of 150 MHz and amplitude of 60 ADC units RMS
has been used, added to a white noise with a RMS level of 0.72 ADC units. 
In this way the ADC operates in a noise-like regime, and 
the quantization process is very linear. The resulting channelized signal 
has been integrated for 4 seconds, using the on-board spectrometer, 
and the spectrum of the noise has then been subtracted. The resulting 
spectrum is shown in figure \ref{fig:response2}, both before and after 
the subtraction of the white noise spectrum, measured separately. 
The difference spectrum shows
harmonics of the original tone, due to distortion in the analog amplifier,
but not other spurious signal up to about -88 dB. 
No strong harmonics are present with low level input signals, or with the
internal digital sinewave generator.
The resulting noise floor 
is still dominated by compression in the analog amplifier, that prevents an
accurate subtraction of the noise, and by the limited dynamic range (32 bits) 
in the spectrometer readout. Further tests, with higher noise level, 
lower tone amplitude and much longer integration times, will be used to 
validate the expected -95 dB stopband rejection. 

Quantization noise introduced by the design has been estimated using
a VHDL simulation of the actual design. The simulation test signal
is a Gaussian white noise of 26 ADC units RMS, and produces a flat
channelized signal with a uniform RMS amplitude of 766 units.  The VHDL
model output has then been compared to an ideal floating point model of
the filterbank.  The total quantization noise has a RMS value of 5.2 units, 
i.e. an excess noise power of $4.5\,10^{-5}$.
80\% of the noise power is due to the polyphase filter and 20\% to the FFT.
Considering an input RMS signal level comprised between 5 and 32 ADC units
(the range required for a good linearity of the digitization process),
the excess noise introduced by the channelizer is at most 0.12\% of the
signal intrinsic noise, and about $1/3$ of the added noise in the 8 bit ADC
converter.

\subsection{Beamformer performance}

\begin{figure}
\begin{center}
\includegraphics[width=0.9\textwidth]{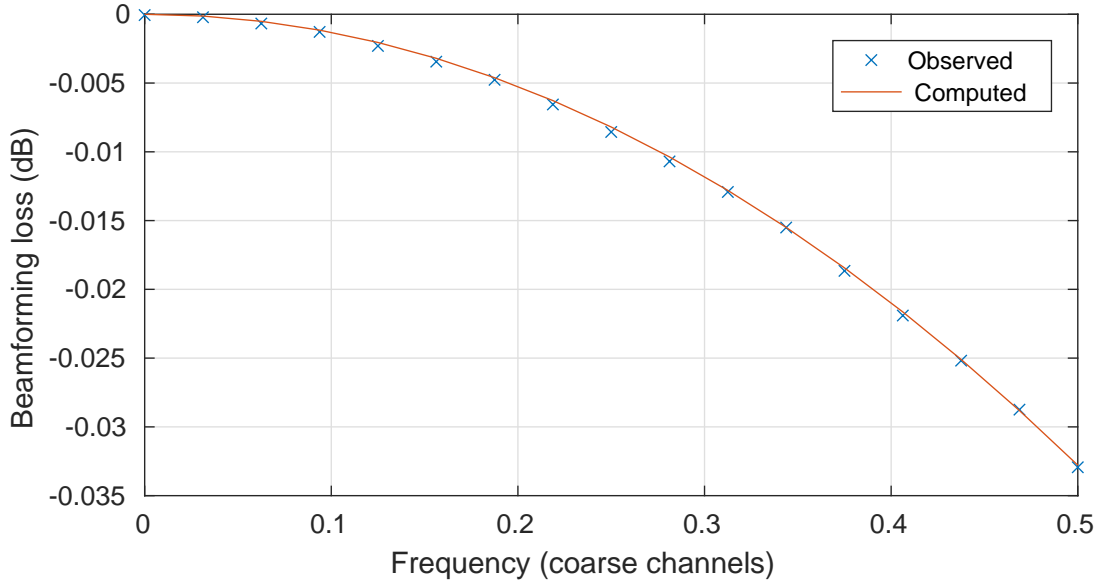}
\end{center}
\caption{Measured beamforming loss with respect to an ideal beamformer, as
a function of the offset from the channel center}
\label{fig:beamform_loss}
\end{figure}

The beamformer performance has been evaluated using a digital 
sinewave generator. The same sinewave is sent in parallel to 
all the tile inputs, using the input delay
compensation stage to simulate a range of physical delays in the antenna
signals and correcting them in the frequency domain. The resulting beamformed 
sinewave is compared to the sinewave produced by a single antenna, 
multiplied by the number of beamformed inputs. 
Simulated delays are in the range of $\pm 40$ ADC samples, corresponding
to a 38m diameter station observing at an elevation of 40 degrees,
i.e. somewhat larger than those expected for a LFAA station. 
The simulation result is shown in figure \ref{fig:beamform_loss}. 
The graph shows the beamforming loss as a function
of the frequency offset of the tone with respect to the frequency
channel center. Frequency domain beamforming is correct only 
for that frequency, and beamforming losses increase quadratically 
up to 0.033 dB at the channel edges. For comparison, the theoric loss
for this architecture is also shown. 

The station beamformer has been verified using data patterns. When the
same data pattern is generated in all the TPM belonging to a station,
the resulting pattern at the end of the TPM chain consists of the
starting pattern multiplied by the number of TPMs in the chain. The
station data stream can be directed to a workstation where it can be
acquired and verified. Alternatively, it can be transmitted back to
the first TPM where a checker verifies the consistency of the data
against the same pattern generated locally and multiplied by the
number of TPM in the chain. In this way, using different types of
patterns, it is possible to verify the operation of the station
beamformer continuously for long period of time.

\section{Conclusions}

The firmware in the Tile Processing Module for the SKA Low Frequency
Aperture Array implements a frequency domain beamformer, and the first
stage of a high resolution spectroscopic channelizer. It allows for
up to 8 independent simultaneous beams, with a total bandwidth of 300
MHz. Its innovative design for an oversampling channelizer allows for a
seamless reconstruction of the observed band, and can be easily adopted
in any channelization system with a very large (up to $~10^7$) number
of channels. The distributed beamformer architecture allows for a dynamic
reconfiguration of the antennas composing each station.

The design has been optimized for resource usage. In particular the filter
performance is limited by the number of available hard multipliers in
the adopted FPGA (Xilinx Kintex Ultrascale XCKU040). Fine tuning of the
design will be required to reduce the power usage, a critical resource
for the SKA telescope. We are looking to an alternative FPGA with more
usable resources, reduced power consumption, or both.

The design has been tested by simulations, with the ADC, channelizer and
tile beamformer already tested in the hardware. A first version of the
complete design will be used as part of the Aperture Array Verification
System demonstrator, in the first half of 2017 which will see the
deployment of 400 antennas on the LFAA site.

Testing of the beamformer and digital firmware design will be carried
out on both astronomical and artificial sources in order to validate
the electromagnetic models of the station beam patterns. Should these
tests be successful we envisage that the firmware will be deployed in
the SKA-1 LOW system which will see the installation of 8192 of these
digital boards capable of beamforming a total of 131072 antennas into 512
stations, before being correlated in a Central Signal Processing building.

The system will also be used at Sardinia Radio Telescope (SRT) site (Cagliari, 
Italy) as a channelizer/beamformer for the Sardinia Array Demonstrator (SAD) 
telescope, a system composed of 128 dual polarization Vivaldi antennas 
operating between 270 and 420 MHz, and in a beamformer for a phased array feed.  

\bibliographystyle{ws-jai}

\bibliography{JAI_TPM_FW}

\end{document}